\documentclass[a4paper,12pt]{article}
\usepackage[utf8]{inputenc}
\usepackage[T1]{fontenc}
\usepackage[english]{babel}
\usepackage{color}
\usepackage{amssymb}
\usepackage{amsmath}
\usepackage{ae}
\usepackage{slashed}
\usepackage{graphics}
\usepackage{graphicx}
\usepackage{epstopdf}
\usepackage{url}

\newcommand{\g}{\gamma}
\newcommand{\p}{\partial}
\newcommand{\lb}{\bar{\lambda}}
\newcommand{\la}{\lambda}

\newcommand{\zb}{\bar{z}}
\newcommand{\ib}{\bar{\imath}}
\newcommand{\jb}{\bar{\jmath}}
\newcommand{\ls}{\left[}
\newcommand{\lp}{\left(}
\newcommand{\rs}{\right]}
\newcommand{\rp}{\right)}
\newcommand{\K}{K\"ahler}
\newcommand{\be}{\begin{equation}}
\newcommand{\ee}{\end{equation}}
\newcommand{\bea}{\begin{align}}
\newcommand{\eea}{\end{align}}

\usepackage{cite}
\usepackage{epsfig}
\usepackage{times}
\usepackage[small]{caption}
\usepackage{ifthen}
\usepackage{longtable}
\usepackage{multirow}
\numberwithin{equation}{section}
\newcommand{\rf}[1]{(\ref{#1})}
\newcommand{\bbox}{\lower.2ex\hbox{$\Box$}}

\usepackage[colorlinks=true,linktoc=page,linkcolor=blue,citecolor=blue,filecolor=blue,urlcolor=blue]{hyperref}
\usepackage{pstricks}

\usepackage{color}
\definecolor{darkgreen}{rgb}{0,.5,0}

\begin{document}
\selectlanguage{english}
\setcounter{secnumdepth}{3}
\frenchspacing
\pagenumbering{roman}

\null{\vspace{\stretch{1}}}
\begin{center}
{\Large {\bf From Linear to Non-linear Supersymmetry\\

\

via Functional Integration}}\\
\vspace{\stretch{1}}
{\normalsize Renata Kallosh, Anna Karlsson and Divyanshu Murli}\\
\vspace{\stretch{1}}
{\small Stanford Institute of Theoretical Physics and Department of Physics,\\ Stanford University, Stanford, CA 94305 USA}\\
\vspace{\stretch{1}}
\end{center}

\begin{abstract}\noindent
We derive a complete pure de Sitter supergravity action with non-linearly realized supersymmetry and its rigid limit, the Volkov--Akulov action, from the corresponding models with linear supersymmetry, by computing the path integral in the limit of infinite sgoldstino mass. In this, we use a non-Gaussian functional integration formula that was recently discovered in a derivation of de Sitter supergravity from the superconformal theory. We also present explicit examples of pure dS supergravity and the case with one matter multiplet. These two simple examples serve as a test and a demo of the universal action formula valid for de Sitter supergravities with general matter coupling.
\end{abstract}
\vspace{\stretch{2}}
\noindent\makebox[\linewidth]{\rule{\textwidth}{0.4pt}}
{\footnotesize \texttt{e-mails: kallosh@stanford.edu, annakarl@stanford.edu, divyansh@stanford.edu}}
\thispagestyle{empty}
\newpage
\pagenumbering{arabic}
\tableofcontents{}
\section{Introduction}
There are many ways to derive actions with $D=4$, $\mathcal{N}=1$ non-linearly realized supersymmetry, starting with the Volkov--Akulov (VA) goldstino action \cite{Volkov:1973ix,Rocek:1978nb,Ivanov:1978mx,Lindstrom:1979kq,Casalbuoni:1988xh,Komargodski:2009rz,Kuzenko:2010ef}. The interest to these theories is due to their recent applications in cosmology \cite{Antoniadis:2014oya,Ferrara:2014kva,Kallosh:2014via,Dall'Agata:2014oka,Ferrara:2015gta,Dudas:2015eha} and in string theory \cite{Kachru:2003aw,Kallosh:2014wsa,Bergshoeff:2015jxa,Kallosh:2015nia,Bandos:2015xnf}. The term `nilpotent' was introduced in \cite{Ferrara:2014kva} as a short definition of the chiral superfield satisfying the constraint $S^2(x, \theta)=0$.

De Sitter supergravity models with a non-linearly realized local supersymmetry were derived recently from the superconformal models underlying supergravity when the nilpotent constraint on a chiral multiplet was imposed \cite{Bergshoeff:2015tra,Hasegawa:2015bza,Kallosh:2015sea}, or from the off-shell supergravity with the constraint \cite{Kallosh:2015tea,Schillo:2015ssx}. Other methods use the constraint chiral curvature superfield, or are based on a complex linear superfield \cite{Kuzenko:2015yxa}, see for example a review of various approaches in \cite{Dudas:2015eha}. A superspace method to derive an analogous action, including fermions, from brane induced supersymmetry breaking was recently proposed in \cite{Bandos:2015xnf}. The term `de Sitter supergravity' was introduced in \cite{Bergshoeff:2015tra} where it was shown that the coupling of supergravity to a nilpotent multiplet leads to a cosmological constant $\Lambda= f^2- 3 m_{3/2}^2 m_{Pl}^2$, which can be positive. Here $f^2$ is a parameter describing the goldstino coupling, i.e. a vacuum value of the auxiliary field of the nilpotent multiplet, or in the context of a string theory: a tension of the uplifting $\overline {D3}$ brane. The universal negative contribution to the cosmological constant is from the gravitino mass. In the general class of de Sitter supergravities \cite{Kallosh:2015tea,Schillo:2015ssx} there is a universal positive contribution $f^2$, a universal negative contribution $- 3 m_{3/2}^2 m_{Pl}^2$, as well as additional positive model dependent terms. We will often use the term `de Sitter supergravity' to replace a longer expression: supergravity interacting with a nilpotent multiplet.

The purpose of this paper is to clarify the relation between linear and non-linear supersymmetries and, in particular, to derive the non-linear actions from the linear models, where $S(x, \theta)$ is an unconstrained chiral multiplet. Another goal of this paper is to explore the models of de Sitter supergravity which are the most interesting for cosmological applications. Specifically, we will present the explicit de Sitter supergravities with a nilpotent multiplet (pure), and the one with one additional chiral (matter) multiplet, by using a general action formula in condensed notation, valid for an arbitrary number of matter multiplets \cite{Kallosh:2015tea,Schillo:2015ssx}. These can be compared with the actions in \cite{Hasegawa:2015bza}, which were derived using a somewhat different method. In this way we will detail and confirm the general result in \cite{Kallosh:2015tea,Schillo:2015ssx} as well as bring it to a form useful for cosmological and particle physics applications.

For the explicit relation between the models with linear and non-linear supersymmetry, we will use an observation in \cite{Komargodski:2009rz} concerning an approximate relation between the models in question, which we will show is possible to make exact, to all orders. The relation is as follows. In the linear model there is a fundamental scalar, the sgoldstino, and a fermion, the goldstino. In \cite{Komargodski:2009rz}, it was shown how one can start with a linear supersymmetry action and eliminate the sgoldstino on shell assuming that the energies are below the masses of the sgoldstino in the IR limit of the quantum field theory: 
\be
E^2\ll m^2_s.
\label{KS}\ee
In this way one recovers the leading terms of the VA fermionic action without scalars, and the process also makes it possible to see why there is only a fermion left in the action at low energies, i.e. below the mass scale of the scalar field sgoldstino.

This procedure may be developed in two ways, as we will show: 
\begin{enumerate}
\item A complete VA action \cite{Volkov:1973ix,Komargodski:2009rz,Kuzenko:2010ef} can be obtained from the linear model by requiring the corresponding path integral to exist in the limit when the mass of the sgoldstino is infinite:
 \be
 m^2_s \rightarrow \infty,
\label{us}\ee
without any bound on the energies, such as in \rf{KS}. Importantly, that very bound at fixed $m_s$ prevents a consistent derivation of the higher derivative terms required in the VA action to preserve the non-linearly realized supersymmetry.

\item The same method of computing the path integral in the limit \rf{us} allows for a derivation of the complete pure non-linear de Sitter supergravity \cite{Bergshoeff:2015tra,Hasegawa:2015bza} from the standard linear supergravity model with one chiral multiplet.
\end{enumerate}

Concerning our second goal, i.e. to present explicit dS supergravity models based on \cite{Kallosh:2015tea}, we would like to point out the difference between the derivations of dS supergravities in \cite{Bergshoeff:2015tra,Kallosh:2015tea} and \cite{Hasegawa:2015bza}. Both approaches begin with an off-shell action where the auxiliary field $F$ of the nilpotent chiral multiplet $(z, \chi, F)$ is present. However, the scalar field $z(x)$ is not a fundamental field, but a bilinear function of the spinors in the sgoldstino:
\be
z= {(\chi)^2\over 2 F}
\ee
This appearance of $F$ in the inverse power makes standard Gaussian integration useless, and an extra effort is required for a derivation of the action for the physical fields. There are different ways to do this. In \cite{Hasegawa:2015bza}, the functional integration of the auxiliary fields in the path integral is performed directly, taking into account the positive and negative powers of $F$ in the action. The explicit expressions for the pure dS case and dS with a matter multiplet were derived. However, this direct method becomes more and more complicated with more general couplings. On the other hand, in \cite{Kallosh:2015tea,Schillo:2015ssx} the integration of the auxiliary fields is performed in a universal way for general couplings following a proposal in \cite{Bergshoeff:2015tra,Kallosh:2015sea}, based on certain features of the nilpotent multiplets. For example, the auxiliary field $F$ can be shifted by a function linear or quadratic in $\chi$ to $F'= F+ {\cal O}(\chi)$. Expressions like
\be
{(\chi)^2\over F}= {(\chi)^2\over F'} 
\ee
remain invariant under this shift since $\chi^3=0$. The existence of the shift symmetry made it possible to argue for that instead of solving a complicated, non-universal equation of motion for $F$ (especially in matter coupled models) one can solve the equation for a shifted auxiliary field $F'$. The subsequent equation has a universal solution, which gives a universal form of the action, presented in \cite{Kallosh:2015tea} in condensed notation. Here, we will extract the explicit expressions for the complete actions in the two previously mentioned cases, which might be useful for various applications. 

The paper is organized as follows. In Sec. 2 we review the result of the non-Gaussian integration of the auxiliary field $F$, as derived in \cite{Bergshoeff:2015tra}. It will be used as a tool for the derivation of the non-linear action from the linear supersymmetry and supergravity models, which we present in Sec. 3. In Sec. 4, we review the general coupling to supergravities with a nilpotent multiplet, before presenting an explicit pure dS supergravity as the first example of the general formula in \cite{Kallosh:2015tea}. It is different from the action in \cite{Bergshoeff:2015tra} due to a difference in the choice of local Weyl symmetry, R-symmetry and special supersymmetry gauge fixing, when starting with superconformal theory. We also present the dS supergravity action coupled to a nilpotent and a matter multiplet. In Sec. 5, the discussion, we summarize our results and suggest possible directions for future work.

\section{Non-Gaussian integration of the auxiliary field of the nilpotent multiplet}\label{s.nGInt}
We begin with a reminder of the simple, well-known procedure for eliminating auxiliary fields by Gaussian integration. The action is quadratic and linear in auxiliary fields, which can be integrated out
\be
S_G\equiv \bar F F - F\bar f - \bar F f \qquad \Rightarrow \qquad \int dF d\bar F e^{iS_G}\propto e^{- i\bar f f}.
\ee
The result is the original action taken at the on-shell Gaussian value of the auxiliary field
\be
F_{\rm on-shell}= f,
\ee
\be
S_{\rm on-shell}= S_G|_{F= f}= -f\bar f.
\ee

Now, we consider the case when the action has a particular non-Gaussian dependence on auxiliary fields, which allows for their elimination, on shell. The action is
\be
S_{nG}\equiv \bar F F - F\bar f - \bar F f + \bar z A z +\bar z B + z\bar B+C\big|_{z={(\chi)^2\over 2 F}}
\label{nG}\ee
Here, $z$ is the first component of the nilpotent chiral multiplet $Z(x, \theta)$:
\be
Z^2(x, \theta)=0 \qquad \Rightarrow \qquad z={(\chi)^2\over 2 F}: \qquad z\chi = 0, \qquad z^2=0,
\ee
with the notation such that
\be
(\chi )^2 \equiv \bar \chi P_L\chi, \qquad z\chi=zP_L\chi.
\ee
The closed form answer for this action is, as shown in \cite{Bergshoeff:2015tra},
\be
S_{\rm on-shell}= \Big [S_{nG}- \Big |{\bar z (A z+B)\over F} \Big|^2 \Big] _{F=f},
\label{result}\ee
where the on-shell value of the auxiliary field was found to be
\be
F_{\rm on-shell}= f \Big[ 1+ {\cal A} \Big( 1- 3 \bar {\cal A} + {(\chi)^2\over 2 f^2 \bar f} \bar B\Big) \Big].
\ee
The corresponding on-shell value of the field $z$ is
\be
z_{\rm on-shell}= {(\chi)^2\over 2F}= {(\chi)^2\over 2f} (1-{\cal A})
\ee
where
\begin{equation}\label{defcalA}
{\cal A}= \frac{ (\bar \chi)^2}{ 2f \bar f^2} \left(A\, \frac{(\chi)^2}{2f} + B\right)= {1\over f\bar f} \Big (\bar z (Az+B)\Big)\Big|_{z= {(\chi)^2\over 2f}} \,.
\end{equation}
It follows that whenever we have an action like \rf{nG}, functional integration can be performed, with the result as given in \rf{result}. Moreover, this formula explains why the fermion action in \cite{Bergshoeff:2015tra} is complete and why it is more difficult to get the complete fermion action in a closed form through other approaches. In particular, in the case of global supersymmetry we have $A=\Box$ and $B=0$, so that $S_{\rm on-shell}$ obtained by this method, starting from 
$S_{nG}$, is given by the VA action \cite{Volkov:1973ix,Komargodski:2009rz,Kuzenko:2010ef}
\begin{equation}
e^{-1}\mathcal{L}=-|f|^2+\frac{(\bar \chi)^2}{2\bar{f}}\Box\frac{(\chi)^2}{2f} -\frac{1}{2}\bar{\chi}\slashed{\nabla}\chi-\frac{(\chi)^2 (\bar \chi)^2}{16|f|^4} \left|f^{-1}\Box (\chi)^2\right|.
\label{VA}\end{equation}

\section{From linear to non-linear supersymmetry in examples, via non-Gaussian integration}\label{s.fL2nLSUSY}
Our general strategy is the following. We generalize the Komargodski--Seiberg construction \cite{Komargodski:2009rz} to all energy scales with the purpose to reproduce the full VA Lagrangian. Specifically, we take the scalar field $z$ in the chiral multiplet 
\be
\mathbf{ Z}(x, \theta) = z + \theta \chi + \theta^2 F
\ee
 to be extremely heavy, which we will show to be equivalent to imposing the nilpotency condition. We take
\begin{gather}\begin{aligned}\label{eq.cKW1}
K(z,\zb)&=z\zb-{c\over M^2} z^2 \zb^2:\quad c\in\mathbb{R},\quad c>0\\
W(z)&=f z
\end{aligned}\end{gather}
It is easy to see that the mass of the sgoldstino field $z(x)$ is
\be
m^2_{z} = 4c {f^2\over M^2},
\ee
which means that for $c\rightarrow\infty$, the sgoldstino is infinitely heavy. Here, we would like to find out if such a limit of the path integral for the linear model defined in \rf{eq.cKW1} exists, under certain conditions. In fact, we will establish that such a limit exists in both the case of a global and a local supersymmetry, and that both cases, unsurprisingly, allow for an alternative to original derivation of the VA global and pure dS local actions, complete to all orders in fermionic terms.

To identify the condition for the existence of the limit $c\rightarrow\infty$ the path integral has to be studied off-shell, when linear supersymmetry also is realized off-shell. In the setting of global linear supersymmetry this means
\be
\int dF d\bar F d z d \bar z d\chi d \bar \chi e^{iS (z, \bar z, \chi, \bar \chi, F, \bar F)}.
\label{pathInt}\ee
The component action is constructed in a standard way, given the corresponding \K\, potential and superpotential. In superfield form it is given by
\be
{\cal L} = \int d^4 \theta\Big ( \mathbf{ Z} \mathbf{ \bar Z} - {c\over M^2} (\mathbf{ Z} \mathbf{ \bar Z})^2 \Big) + \int d^2\theta \big[f \mathbf{ Z} +\rm{h.c.}\big]
\label{super}\ee
This form is convenient as it shows that the complete action has at most a linear dependence on $c$:
\be
{\cal L} (c)= {\cal L}|_{c=0} + c \, {\cal L}^1.
\ee
We will therefore, in the case of global supersymmetry, require
\be
{\cal L}^1 =0
\label{cond}\ee
as a condition for the existence of the limit $c\rightarrow\infty$.

In the case of local supersymmetry we find that all higher powers of $c$ are present in the action, for example from the terms with $e^K\sim e^{-{c\over M^2} z^2 \zb^2} $ and other sources:
\be
{\cal L}_{loc}(c) = {\cal L}_{loc}|_{c=0} + c \, {\cal L}^1_{loc} +\sum_{n\geq2} c^n {\cal L}^n_{loc}.
\label{loc}\ee
However, once we require that 
\be
{\cal L}^1_{loc} =0
\label{cond1}\ee
an analysis shows that all higher order terms in $c$ also vanish, so that the limit exists. Consequently, what remains is to perform the functional integration of the original integral with the originally linearly realized supersymmetry with the action in \rf{pathInt} under conditions \rf{cond}, \rf{cond1} respectively, for global and local supersymmetry, and find out the result of the integration.

Since the \K\, potential is the same for both the global and the local case, we will begin by studying the \K\, geometry, where we will denote the corresponding linear supermultiplet by $(z^1,\chi^1,F^1)$ for convenience of dealing with holomorphic and anti-holomorphic indices. Here, $\chi^1$ is a Majorana spinor which in the rest of this article will be split into its Weyl components
\begin{equation}\label{eq.chi1b1}
\chi^1=P_L\chi^1,\qquad \chi^{\bar{1}}=P_R\chi^1: \qquad (\chi^1)^2=\bar{\chi}^1\chi^1,\qquad (\chi^{\bar{1}})^2=\bar{\chi}^{\bar{1}}\chi^{\bar{1}}.
\end{equation} 
In this setting, the K\"ahler potential and the superpotential is
\begin{gather}\begin{aligned}\label{eq.cKW}
K(z^1,\zb^{\bar{1}})&=z^1\zb^{\bar{1}}-{c\over M^2} (z^1)^2(\zb^{\bar{1}})^2: \quad c>0\\
W(z^1)&=fz^1,
\end{aligned}\end{gather}
and we observe that the metric and its inverse are
\begin{subequations}
\begin{equation}
g_{1\bar{1}} = 1 - \frac{4c}{M^2} z^1 \zb^{\bar{1}} \quad\Rightarrow\quad g^{1\bar{1}}=\frac{1}{g_{1\bar{1}}},
\end{equation} 
while the non-vanishing Christoffel symbol ($\Gamma^\alpha_{\beta\g}=g^{\alpha\bar{\delta}}\p_{\beta}g_{\g\bar{\delta}}$) and Riemann tensor components are
\begin{gather}\begin{aligned}
\Gamma_{11}^1&=-\frac{4c}{M^2}g^{1\bar{1}}\zb^{\bar{1}},\\
\Gamma_{\bar{1}\bar{1}}^{\bar{1}}&=-\frac{4c}{M^2}g^{1\bar{1}}z^1,\\
R_{1\bar{1}1\bar{1}}&=-\frac{4c}{M^2}-\frac{16c^2}{M^4}g^{1\bar{1}}z^1\zb^{\bar{1}}.
\end{aligned}\end{gather}
\end{subequations}

\subsection{Global supersymmetry}
The standard $\mathcal{N} =1$ supersymmetric Lagrangian with chiral multiplets and without gauge multiplets is given by \cite{Freedman:2012zz}:
\begin{gather}\begin{aligned}
e^{-1}\mathcal{L}=&g_{\alpha\bar{\beta}}\bigg[-\p_\mu z^\alpha\p^\mu z^{\bar{\beta}}-\frac{1}{2}\bar{\chi}^\alpha P_L\slashed{\nabla}\chi^{\bar{\beta}}-\frac{1}{2}\bar{\chi}^{\bar{\beta}} P_R\slashed{\nabla}\chi^{\alpha}+\\
&+(F^\alpha-\frac{1}{2}\Gamma^\alpha_{\beta\g}\bar{\chi}^\g P_L\chi^\beta)(F^{\bar{\beta}}-\frac{1}{2}\Gamma^{\bar{\beta}}_{\bar{\g}\bar{\alpha}}\bar{\chi}^{\bar{\g}} P_L\chi^{\bar{\alpha}})\bigg]+\\
&+\frac{1}{4}R_{\alpha\bar{\gamma} \beta \bar{\delta}}\bar{\chi}^\alpha P_L\chi^\beta\bar{\chi}^{\bar{\g}}P_R \chi^{\bar{\delta}}+\\
&+\left[W_\alpha F^\alpha-\frac{1}{2}W_{\alpha\beta}\bar{\chi}^\alpha P_L\chi^\beta+\rm{h.c.}\right].
\end{aligned}\end{gather}
With the K\"ahler and superpotential in \eqref{eq.cKW}, using notation in \eqref{eq.chi1b1}, this Lagrangian is
\begin{gather}\begin{aligned}
e^{-1}\mathcal{L}=&g_{1\bar{1}}\Big[-\p_\mu z^1\p^\mu \zb^{\bar{1}}- \frac{1}{2} \bar{\chi}^1 P_L \gamma^{\mu} \nabla_{\mu} \chi^{\bar{1}} - \frac{1}{2} \bar{\chi}^{\bar{1}} P_R \gamma^{\mu} \nabla_{\mu} \chi^1+\\
&+(F^1-\frac{1}{2}\Gamma^1_{11}(\chi^1)^2)(\bar{F}^{\bar 1}-\frac{1}{2}\Gamma^{\bar{1}}_{\bar{1}\bar{1}}(\chi^{\bar{1}})^2)\Big]+\frac{1}{4}R_{1\bar{1}1\bar{1}}(\chi^1)^2(\chi^{\bar{1}})^2+fF^1+\bar{f}\bar{F}^{\bar 1}=\\
=& z^1 \Box \zb^{\bar{1}} - \frac{1}{2} \bar{\chi}^1 P_L \gamma^{\mu} \nabla_{\mu} \chi^{\bar{1}} - \frac{1}{2} \bar{\chi}^{\bar{1}} P_R \gamma^{\mu} \nabla_{\mu} \chi^1 + F^1 \bar{F}^{\bar 1} + fF^1 + \bar{f} \bar{F}^{\bar 1}-\\
 &- \frac{c}{M^2} | 2z^1F^1 - (\chi^1)^2 |^2 + \mathcal{O}(\chi^1z^1)+ \mathcal{O}(\chi^{\bar{1}}\zb^{\bar{1}})
\end{aligned}\end{gather}
The terms in the action which seem to have a quadratic dependence on $c$ in the moduli space curvature and the ones quadratic in $\Gamma$ actually cancel, in agreement with \rf{super}. We find that the condition of existence of the limit $c\rightarrow\infty$ in \rf{cond} becomes
\be
{\cal L}^1 =- \frac{c}{M^2} | 2z^1F^1 - (\chi^1)^2 |^2 + \mathcal{O}(\chi^1z^1)+ \mathcal{O}(\chi^{\bar{1}}\zb^{\bar{1}})
= 0,
\label{condE}\ee
which we solve by requiring that 
\be \label{eq.z1cond}
z^1 = \frac{(\chi^1)^2}{2F^1}.
\ee
This is a necessary condition to eliminate the first term. In addition, the property $z^1\chi^1=0$ means that the $\mathcal{O}(\chi^1z^1)+ \mathcal{O}(\chi^{\bar{1}}\zb^{\bar{1}})$ vanish as well. 

Thus in the limit $m^2_z \rightarrow \infty$, we obtain the same constraint \eqref{eq.z1cond} as \cite{Komargodski:2009rz}, but without the restrictions on the energy scale in \rf{KS}, since we study the linear supersymmetry model at $m^2_z \rightarrow \infty$ and $E^2\ll \infty$ with the purpose to derive a complete VA action. The limiting Lagrangian becomes
\begin{equation}
e^{-1}\mathcal{L} = z^1 \Box \zb^{\bar{1}} - \frac{1}{2} \bar{\chi}^1 P_L \gamma^{\mu} \nabla_{\mu} \chi^{\bar{1}} - \frac{1}{2} \bar{\chi}^{\bar{1}} P_R \gamma^{\mu} \nabla_{\mu} \chi^1 + F^1 \bar{F}^{\bar 1} + fF^1 + \bar{f} \bar{F}^{\bar 1}.
\end{equation}

We can now simply use the non-Gaussian integration trick employed in Sec. \ref{s.nGInt} to eliminate the auxiliary field $F^1$. Upon integrating out $F^1$ the action \rf{result} becomes equal to the complete VA action \rf{VA}. However, note that in comparison with this equation, and \eqref{nG}, $f^1=-\bar{f}$, where $f^1$ is the on-shell value of $F^1$. This relation between $f^1$ and the superpotential is a consequence of the form of the action.

\subsection{Local supersymmetry}
In the case of local supersymmetry, we focus again on an example of supergravity interacting with one linear, unconstrained chiral multiplet $(z^1,\chi^1,F^1)$ with the \K\, potential and superpotential given in \rf{eq.cKW}. We claim that in the limit $c\rightarrow\infty$, performed using the corresponding local supersymmetry action, we will be able to derive the pure dS supergravity, as already indicated in the beginning of this section. We start with the off-shell supergravity action, following \cite{Kallosh:2015tea} in the case of a single unconstrained chiral multiplet:
\begin{equation}
e^{-1}\mathcal{L}=(F^1-F_G^1)g_{1\bar 1}(\bar F^{\bar{1}}-\bar F_G^{\bar 1})+e^{-1}\mathcal{L}_{\text{book}},
\end{equation}
where the Gaussian solution is
\be
F_G^{\alpha}=-e^{\frac{K}{2}}g^{\alpha\bar{\beta}}\overline{\nabla}_{\bar{\beta}}\overline{W}+\frac{1}{2}\Gamma^\alpha_{\beta\g}\bar{\chi}^\beta\chi^\g.
\ee
With \eqref{eq.z1cond}, the extra terms arising in the Lagrange multiplier, scalar potential, the fermion mass Lagrangian and the four fermion Lagrangian due to the extra term in the K\"ahler potential proportional to $c$, cancel. The first gives a $c$-dependent contribution (disregarding terms with $\chi^1 z^1$)
\begin{subequations}
\begin{equation}
-\frac{4c}{M^2}z^1\zb^{\bar{1}}F^1\bar{F}^{\bar 1}-\frac{4c}{M^2}|f^1|^2+\frac{2c}{M^2}\left((F^1+f^1)z^1(\chi^{\bar{1}})^2+(\bar{F}^{\bar 1} +\bar{f}^{\bar{1}})\zb^{\bar{1}}(\chi^1)^2\right),
\end{equation}
\begin{equation}
f^1= -{1\over g_{1\bar{1}}} e^{\frac{K}{2}} \bar f\Big|_{c=0},
\end{equation}
\end{subequations}
apart from the expression present with $c=0$. The second of these terms gets cancelled by the contribution from the scalar potential (as it does also with $c=0$), and the other terms with $f^1$ by the contribution from the fermion mass Lagrangian. The rest of the terms pair up with the four fermion Lagrangian contribution (from the Riemann tensor) into:
 \begin{equation}
 -\frac{c}{M^2}\left|2z^1F^1-(\chi^1)^2\right|^2
 \end{equation}
We therefore find that with \eqref{eq.z1cond}, and taking into account all other $c$-dependent terms, the limit $c\rightarrow\infty$ is well-defined and the action takes the form shown in \rf{nG}. Integrating the auxiliary field $F^1$ using the non-Gaussian formula \rf{result} we find the pure dS supergravity action in the form given in \cite{Kallosh:2015tea}
\begin{equation}
e^{-1}\mathcal{L}\Big |_{z^1=\frac{(\chi^1)^2}{2F^1}} \quad \xrightarrow{c\rightarrow\infty} \quad e^{-1}\mathcal{L}_{\text{Pure dS}}.
\end{equation}
We will present an explicit expression for this action below.

\section{General Matter-coupled de Sitter supergravity}
The most general case of a K\"ahler potential $K$, a superpotential $W$ and a holomorphic gauge kinetic function $f_{AB}$ in the case when one chiral multiplet $z^1$ is nilpotent can be presented as follows:
\begin{eqnarray}
K(z^1,\zb^{\bar{1}},z^i,\zb^{\ib})&=&K_0(z^i,\zb^{\ib})+z^1K_1(z^i,\zb^{\ib})+\zb^{\bar{1}}K_{\bar{1}}(z^i,\zb^{\ib})+z^1\zb^{\bar{1}}g_{1\bar{1}}(z^i,\zb^{\ib})\nonumber\\
W(z^1,z^i)&=&W_0(z^i)+z^1W_1(z^i)\equiv g(z^i)+z^1f(z^i)\\
f_{AB}(z^1,z^i)&=&f_{AB0}(z^i)+z^1f_{AB1}(z^i)\nonumber
\end{eqnarray}
Here we have taken into account that all dependence on $z^1$ and $\zb^{\bar{1}}$ is restricted by the condition $(z^1)^2=0$ and $(\zb^{\bar{1}})^2=0$.
In \cite{Kallosh:2015tea} a simplifying assumption\footnote{A complete action without this simplifying assumption was constructed in \cite{Schillo:2015ssx}.} was made by requiring that there were no terms in the \K\, potential linear in $z^1$ ($\zb^{\bar{1}}$), i.e. 
\begin{equation}
K_{1}(z^i,\zb^{\ib})=K_{\bar{1}}(z^i,\zb^{\ib})=0.
\label{ass}\end{equation}
Using the assumption \eqref{ass}, the supergravity Lagrangian for a nilpotent chiral superfield, coupled to an arbitrary number of chiral and vector multiplets ($i=2,\ldots, n$), is given in \cite{Kallosh:2015tea}
\begin{subequations}
\begin{equation}\label{eq.L}
e^{-1}\mathcal{L}=\big[e^{-1}\mathcal{L}^{\text{book}}\big]_{z^1=\frac{(\chi^1)^2}{2f^1}}-\frac{(\chi^1)^2(\chi^{\bar{1}})^2}{4g_{1\bar{1}}(f^1\bar{f}^1)^2}\bigg|g_{1\bar{1}}\frac{\Box(\chi^1)^2}{2f^1}+b^1\bigg|^2,
\end{equation}
\begin{equation}
f^1\equiv\frac{1}{g_{1\bar{1}}}\Big(-e^{\frac{K_0}{2}}\bar{W}_{\bar{1}}+\frac{1}{4}\bar{f}_{AB\bar{1}} \lb^AP_R\la^B\Big),
\end{equation}
\begin{gather}\begin{aligned}
b^1\equiv& e^{-1}\frac{\delta\mathcal{L}^\text{book}}{\delta\zb^{\bar{1}}}\bigg|_{\chi^1=\chi^{\bar{1}}=0}=\\
=& e^{K_0}\ls 2 g \bar f +(K_{0,i\jb})^{-1} \lp D_{0,i}g\rp\lp \frac{\partial_{\jb} g_{1\bar 1}}{g_{1\bar 1}}\bar{f} -\bar{D}_{0,\jb} \bar{f}\rp\rs-\cr
&- \frac12 (\bar \chi^{\jb} \slashed{\partial}\chi^1) \ls 3 \partial_{\jb} g_{1 \bar 1} +(\bar{z}^{\bar k} \partial_{\bar k} -z^k \partial_k) \partial_{\jb} g_{1\bar1} \rs+\cr
&+\frac12 e^{\frac{K_0}{2}}\bar{f} \bar \psi_\mu P_L \gamma^{\mu\nu} \psi_\nu - \frac{1}{\sqrt{2}} \bar \psi_\mu \gamma^\nu \gamma^\mu \partial_\nu \chi^1 \big( g_{1 \bar 1} +(\partial_{\jb} g_{1\bar 1}) \bar z^{\jb}\big)+\cr
&+\frac14 e^{\frac{K_0}{2}} f_{ABj} \ \bar \lambda^A P_L \lambda^B \ (K_{0,j\bar k})^{-1} \lp(\partial_{\bar k} +K_{0,\bar k}) -\frac{\partial_{\bar k} g_{1 \bar 1}}{g_{1 \bar 1}} \rp \bar{f} -\cr
&-\frac12e^{\frac{K_0}{2}} \ \bar{\chi}^{\ib} \chi^{\jb} \ \lp \bar{D}_{0,\ib} \bar{D}_{0,\jb} -\frac{\partial_{\ib} \partial_{\jb} g_{1 \bar 1}}{g_{1\bar1}}+\frac{K_{0,k\ib\jb}}{K_{0,k\bar l}} \lp \frac{(\partial_{\bar l} g_{1 \bar 1})}{g_{1\bar1}}-\bar{D}_{0,\bar l}\rp \rp \bar{f}+\cr
&+\frac{1}{\sqrt{2}} e^{\frac{K_0}{2}} \bar{D}_{0,\ib} \bar{f} \ \bar \chi^{\ib}\, \gamma\cdot\psi \, ,
\end{aligned}\end{gather}
\end{subequations}
where $D_{0,i} = \partial_i + K_{0,i}$ and $\bar{D}_{0,\ib} = \partial_{\ib} + K_{0,\ib}$. Here, the $b^1$ can be read off from the action given in \cite{Freedman:2012zz}. The first line of it comes from the scalar potential $V$, and the second line from the kinetic terms for the fermions. The third, fourth and fifth lines come from the fermion mass Lagrangian, except for the second term of the third line, which is of the form $\bar{\psi}_\mu\mathcal{J}^\mu$. The last line contains a term mixing gravitino and spin 1/2 fermions.

It might be interesting to note that a class of models called `Nilpotent Supergravity' in \cite{Dudas:2015eha} allows only a particular choice of the K\"ahler potential:
\begin{equation}
K(z^1,\zb^{\bar{1}},z^i,\zb^{\ib})=K_0(z^i,\zb^{\ib})-3(z^1+\zb^{\bar{1}})e^{\frac{1}{3}K_0(z^i,\zb^{\ib})}+3z^1\zb^{\bar{1}}e^{\frac{2}{3}K_0(z^i,\zb^{\ib})},
\end{equation}
characterized by a single function $K_0(z^i,\zb^{\ib})$. These models have a nilpotent multiplet dual to a chiral curvature multiplet. Meanwhile, starting with the superconformal approach, as proposed in \cite{Kallosh:2015sea} and implemented in \cite{Kallosh:2015tea,Schillo:2015ssx}, there are no such constraints. All of the three functions that appear in the K\"ahler potential can be independent.

\subsection{Explicit pure de Sitter supergravity}
Here we have a coupling of supergravity with a single nilpotent multiplet denoted in \cite{Kallosh:2015tea} as $z^1$. For clarity, we will use for its components the notation $(z^1, \chi^1, F^1)$:
\begin{equation}
\left.\begin{array}{ll}
K=z^1\zb^{\bar{1}}&\\\
W=g+fz^1:&g,f\in\mathbb{C}
\end{array}\right\}\quad\Rightarrow\quad f^1=-\bar{f}
\end{equation}
We know from \cite{Kallosh:2015tea} that for the pure de Sitter case
\be\label{eq:B1puredS}
b^1 = 2 g \bar f +\frac12 \bar f \bar \psi_\mu P_L \gamma^{\mu\nu} \psi_\nu- \frac{1}{\sqrt{2}} \bar \psi_\mu \gamma^\nu \gamma^\mu \partial_\nu \chi^1 \,. 
\ee
The full Lagrangian corresponding to \eqref{eq.L} for this model is
\begin{gather}\begin{aligned}\label{eq.LPuredS}
e&^{-1}\mathcal{L}=\frac{1}{2}\Big[R-\bar{\psi}_\mu\g^{\mu\nu\rho}D_\mu\psi_\rho\Big]-\frac{1}{2}\bar{\chi}^1\g^\mu D_\mu\chi^{\bar{1}}-\frac{1}{2}\bar{\chi}^{\bar{1}}\g^\mu D_\mu\chi^1+\\
&+3|g|^2- |f|^2+\Big[\frac{g}{2}\bar{\psi}_\mu P_R\g^{\mu\nu}\psi_\nu+\frac{f}{\sqrt{2}}\bar{\psi}_\mu\g^{\mu}\chi^1+\rm{h.c.}\Big]+\\
&+\frac{1}{2}\mathcal{L}_{SG,T}+\frac{ie^{-1}}{16}\varepsilon^{\mu\nu\rho\sigma}\bar{\psi}_\mu\g_\nu\psi_\rho\bar{\chi}^{\bar{1}}\g_\sigma\chi^1 -\frac{1}{2}\bar{\psi}_\mu\chi^{\bar{1}}\bar{\psi}^\mu\chi^{1}-\frac{1}{8}(\chi^1)^2(\chi^{\bar{1}})^2-\\
&-\bigg[\frac{1}{2^{3/2}f} \bar{\psi}_\mu\g^\nu(\p_\nu(\chi^{\bar{1}})^2)\g^\mu \chi^1+\frac{g\bar{f}(\chi^{\bar{1}})^2}{f}+\\
&\qquad+\frac{\bar{f}(\chi^{\bar{1}})^2}{4f}\bar{\psi}_\mu P_L\g^{\mu\nu}\psi_\nu -\frac{(\chi^1)^2(\chi^{\bar{1}})^2}{2}+\frac{g(\chi^{\bar{1}})^2}{2^{3/2}f}\bar{\psi}_\mu\g^\mu\chi^1+\rm{h.c.}\bigg]+\\
&+\frac{1}{4|f|^2}\bigg[\frac{i}{8}\varepsilon^{\mu\nu\rho\sigma}\bar{\psi_\mu}\g_\nu\psi_\rho\Big((\chi^{\bar{1}})^2\p_\sigma (\chi^1)^2-(\chi^1)^2\p_\sigma(\chi^{\bar{1}})^2\Big)-\p_\mu (\chi^1)^2\p^\mu(\chi^{\bar{1}})^2\bigg]+\\
&+\frac{(\chi^1)^2(\chi^{\bar{1}})^2}{16|f|^2}\Big[8|g|^2+g\bar{\psi}_\mu P_R\g^{\mu\nu}\psi_\nu + \bar{g}\bar{\psi}_\mu P_L\g^{\mu\nu}\psi_\nu\Big]-\\
&-\frac{(\chi^1)^2(\chi^{\bar{1}})^2}{4|f|^4}\bigg|-\frac{\Box(\chi^1)^2}{2\bar{f}}+\left(2g\bar{f}+\frac{1}{2}\bar{f}\bar{\psi}_\mu P_L\g^{\mu\nu}\psi_\nu-\frac{1}{\sqrt{2}} \bar{\psi}_\mu\g^\nu\g^\mu \p_\nu\chi^1\right)\bigg|^2,
\end{aligned}\end{gather}
with the covariant spinor derivative and supergravity torsion part $\mathcal{L}_{SG,T}$ of the 
\begin{subequations}\label{eq.DLSGT}
\begin{align}
D_\mu&=\p_\mu+\frac{1}{4}{\omega_\mu}^{ab}(e)\g_{ab},\\
L_{SG,T}&=-\frac{1}{16}[(\bar{\psi}^\rho\g^\mu\psi^\nu)(\bar{\psi}_\rho\g_\mu\psi_\nu+2\bar{\psi}_\rho\g_\nu\psi_\mu)-4(\bar{\psi}_\mu\g_\nu\psi^\nu)(\bar{\psi}^\mu\g_\nu\psi^\nu)].
\end{align}
\end{subequations}

The three first lines of \eqref{eq.LPuredS} denote terms independent of $(z^1,\zb^{\bar{1}})$, representing the normal Einstein theory. The first line begins with the kinetic terms for the graviton and continues with parts of the corresponding for the gravitino and the chiral multiplets. This is followed by the contributions from the scalar potential, the fermion mass Lagrangian and the terms that mix the gravitino and spin $1/2$ fermions, whereas the third line contains terms with four fermions.

Lines four to five denote the contribution from terms proportional to $z^1: (B\zb^{\bar{1}}+z^1\bar{B})$. The first of these contains the term with $\bar{\psi}_\mu \mathcal{J}^\mu$, proportional to the basic supercurrent of the gauge, and a term from the scalar potential. The second shows the fermion mass Lagrangian and a final term mixing gravitino and spin 1/2 fermions.

The sixth and the seventh lines in \eqref{eq.LPuredS} corresponds to terms with both $z^1$ and $\zb^{\bar{1}}$, from the kinetic terms for the chiral multiplets, the scalar potential and the fermion mass Lagrangian. The final line represents the extra term specified in \eqref{eq.L}. Here, the $b^1$ can be read off.

\subsection{Explicit de Sitter supergravity coupled to one chiral multiplet }
We continue presenting the nilpotent multiplet as $(z^1, \chi^1, F^1)$. The matter multiplet components will be $(z^2, \chi^2, F^2)$. We make the simple choice
\begin{gather}\begin{aligned}
K(z^1,\zb^{\bar{1}},z^2,\zb^{\bar{2}})&=k(z^2,\zb^{\bar{2}})+z^1\zb^{\bar{1}},\\
W(z^1,z^2)&=g(z^2)+z^1f(z^2),
\end{aligned}\end{gather}
with which we explicitly have
{\small
\begin{gather}\begin{aligned}
e^{-1}\mathcal{L}&=\frac{1}{2}\Big[R-\bar{\psi}_\mu\g^{\mu\nu\rho}D_\mu\psi_\rho+\frac{i}{4}\varepsilon^{\mu\nu\rho\sigma}\bar{\psi_\mu}\g_\nu\psi_\rho\Big(k_2\p_\sigma z^2-k_{\bar{2}}\p_\sigma\zb^{\bar{2}}\Big)\Big]-\\
&-k_{2\bar{2}}\p_\mu z^2\p^\mu\zb^{\bar{2}}-\frac{1}{2}\bar{\chi}^1\g^\mu D'_\mu\chi^{\bar{1}}-\frac{1}{2}\bar{\chi}^{\bar{1}}\g^\mu D'_\mu\chi^1-\frac{k_{2\bar{2}}}{2}(\bar{\chi}^2\g^\mu D'_\mu\chi^{\bar{2}})-\\
&-\frac{k_{2\bar{2}}}{2}(\bar{\chi}^{\bar{2}}\g^\mu D'_\mu\chi^2)-\frac{k_{2\bar{2}2}}{2}\bar{\chi}^{\bar{2}}\g^\mu\chi^2\p_\mu z^2-\frac{k_{\bar{2}2\bar{2}}}{2}\bar{\chi}^{2}\g^\mu\chi^{\bar{2}}\p_\mu z^{\bar{2}}+\\
&+3|\mathrm{g}|^2-|\mathrm{f}|^2-\frac{1}{k_{2\bar{2}}}|\nabla_2\mathrm{g}|^2+\bigg[\frac{k_{2\bar{2}}}{\sqrt{2}}\bar{\psi}_\mu\g^\nu(\p_\nu\zb^{\bar{2}})\g^\mu\chi^2+\frac{\mathrm{g}}{2}\bar{\psi}_\mu P_R\g^{\mu\nu}\psi_\nu-\\
&-\nabla_2\mathrm{f}\bar{\chi}^1\chi^2-\frac{1}{2}\nabla_2^2\mathrm{g}(\chi^2)^2+\frac{1}{\sqrt{2}}\bar{\psi}_\mu\g^{\mu}(\chi^1\mathrm{f}+\chi^2\nabla_2\mathrm{g})+\rm{h.c.}\bigg]+\\
&+\frac{1}{2}\mathcal{L}_{SG,T}+\frac{ie^{-1}}{16}\varepsilon^{\mu\nu\rho\sigma}\bar{\psi}_\mu\g_\nu\psi_\rho(\bar{\chi}^{\bar{1}}\g_\sigma\chi^1+k_{2\bar{2}}\bar{\chi}^{\bar{2}}\g_\sigma\chi^2)-\frac{1}{2}\bar{\psi}_\mu\chi^{\bar{1}}\bar{\psi}^\mu\chi^{1}-\\
&-\frac{k_{2\bar{2}}}{2}\bar{\psi}_\mu\chi^{\bar{2}}\bar{\psi}^\mu\chi^{2}-\frac{1}{8}\Big[(\chi^1)^2(\chi^{\bar{1}})^2+2k_{2\bar{2}}(\bar{\chi}^1\chi^{2})(\bar{\chi}^{\bar{1}}\chi^{\bar{2}})+k_{2\bar{2}}(\chi^2)^2(\chi^{\bar{2}})^2\Big]+\\
&+\frac{1}{4}\bigg(k_{2\bar{2}2\bar{2}}-\frac{|k_{2\bar{2}2}|^2}{k_{2\bar{2}}}\bigg)(\chi^2)^2(\chi^{\bar{2}})^2-\\
&-\bigg[\frac{1}{2^{3/2}} \bar{\psi}_\mu\g^\nu(\p_\nu\frac{(\chi^{\bar{1}})^2}{\mathrm{f}})\g^\mu \chi^1+\frac{\mathrm{g}\bar{\mathrm{f}}(\chi^{\bar{1}})^2}{\mathrm{f}}-\frac{(\chi^{\bar{1}})^2}{2\mathrm{f}k_{2\bar{2}}}\nabla_2\mathrm{g}\overline{\nabla_2\mathrm{f}}+\\
&\qquad+\frac{\overline{\mathrm{f}}(\chi^{\bar{1}})^2}{4\mathrm{f}}\bar{\psi}_\mu P_L\g^{\mu\nu}\psi_\nu -\frac{(\chi^1)^2(\chi^{\bar{1}})^2}{2}-\frac{(\chi^{\bar{1}})^2}{4\mathrm{f}}\Big(2\nabla_2\mathrm{g}\bar{\chi}^1\chi^2+\overline{\nabla_2^2\mathrm{f}}(\chi^{\bar{2}})^2\Big)+\\
&\qquad+\frac{\mathrm{g}(\chi^{\bar{1}})^2}{2^{3/2}\mathrm{f}}\bar{\psi}_\mu\g^\mu\chi^1+\frac{(\chi^{\bar{1}})^2}{2^{3/2}\mathrm{f}}(\bar{\chi}^{\bar{2}}\g^\mu\psi_\mu)\overline{\nabla_2\mathrm{f}}+\rm{h.c.}\bigg]+\\
&+\frac{1}{32}\Big(i\varepsilon^{\mu\nu\rho\sigma}\bar{\psi_\mu}\g_\nu\psi_\rho+k_{2\bar{2}}[\bar{\chi}^2\g^\sigma\chi^{\bar{2}}+\bar{\chi}^{\bar{2}}\g^\sigma\chi^2]\Big)\bigg(\frac{(\chi^{\bar{1}})^2}{\rm f}\p_\sigma \frac{(\chi^1)^2}{\rm \bar{f}}-\frac{(\chi^1)^2}{\rm \bar{f}}\p_\sigma\frac{(\chi^{\bar{1}})^2}{\rm f}\bigg)-\\
&-\frac{1}{4}\bigg[\p_\mu \frac{(\chi^1)^2}{\bar{\mathrm{f}}}\bigg]\bigg[\p^\mu\frac{(\chi^{\bar{1}})^2}{\mathrm{f}}\bigg]+\frac{(\chi^1)^2(\chi^{\bar{1}})^2}{4|\mathrm{f}|^2}\bigg[2|\mathrm{g}|^2-\frac{1}{k_{2\bar{2}}}(|\nabla_2 \mathrm{g}|^2+|\nabla_2 \mathrm{f}|^2)\bigg]+\\
&+\frac{(\chi^1)^2(\chi^{\bar{1}})^2}{16|\mathrm{f}|^2}\Big[\sqrt{2}\bar{\psi}_\mu\g^\mu\chi^2\nabla_2\mathrm{g}+\mathrm{g}\bar{\psi}_\mu P_R\g^{\mu\nu}\psi_\nu-(\nabla_2^2\mathrm{g})(\chi^2)^2 + \rm{h.c.}\Big]-\\
&-\frac{(\chi^1)^2(\chi^{\bar{1}})^2}{4|\mathrm{f}|^4}\bigg|-\frac{\Box(\chi^1)^2}{2\bar{\mathrm{f}}}+\bigg(2\mathrm{g}\bar{\mathrm{f}}-\frac{1}{k_{2\bar{2}}}\nabla_2\mathrm{g}\overline{\nabla_2\mathrm{f}}+\frac{1}{2}\bar{\mathrm{f}}\bar{\psi}_\mu P_L\g^{\mu\nu}\psi_\nu-\\
&\qquad\qquad\qquad\quad-\frac{1}{\sqrt{2}}\bar{\psi}_\mu\g^\nu\g^\mu\p_\nu\chi^1-\frac{1}{2}(\overline{\nabla_2^2\mathrm{f}})(\chi^{\bar{2}})^2+\frac{1}{\sqrt{2}}(\bar{\chi}^{\bar{2}}\g^\mu\psi_\mu)\overline{(\nabla_2\mathrm{f})}\bigg)\bigg|^2,
\label{ds+1}\end{aligned}\end{gather}
}
with $D_\mu$ and $\mathcal{L}_{SG,T}$ as in \eqref{eq.DLSGT} and
\begin{subequations}
\begin{align}
\mathrm{g}&=e^{k/2}g, \quad \mathrm{f}=e^{k/2}f,\\
D'_\mu&=D_\mu-\frac{1}{4}\big(k_2\p_\mu z^2-k_{\bar{2}}\p_\mu\zb^{\bar{2}}),\\
\nabla_i&=e^{k/2}\left[\p_i+(\p_i k)\right]e^{-k/2}=\p_i+\frac{1}{2}(\p_ik)=\p_i+\frac{1}{2}k_i,\\
\nabla_2^2&=e^{k/2}\nabla_2\nabla_2e^{-k/2}=\p_2\p_2+k_2\p_2+\frac{k_{22}}{2}+\frac{k_2k_2}{4}-\frac{k_{2\bar{2}2}}{2}\nabla_2.
\end{align}
\end{subequations}
The structure of the presentation follows that of the pure de Sitter case. The eight first lines represent the contribution independent of $(z^1,\bar{z}^{\bar{1}})$, ordered in the same way as previously mentioned. For example, the contribution from the scalar potential is found in the first three terms of the fourth line, and the terms containing four fermions in lines six to eight.

Line nine to eleven is proportional to $z^1$, followed by the terms containing both $z^1$ and $\bar{z}^{\bar{1}}$, where the last two lines correspond to the additional contribution to the book Lagrangian, specified in \eqref{eq.L}.

Note that although this expression is quite complicated, it is in fact just a detailed version of \eqref{eq.L}. In any analysis, it likely most efficient to first look to the short form, and the details there, for e.g. an appropriate choice of gauge. In the absence of terms proportional to $z^1$, the Lagrangian simply is constituted by the eight first lines in \eqref{ds+1}. Also, in contrast to what is true for e.g. the approach in \cite{Hasegawa:2015bza}, the expression can be easily extended to include more matter multiplets. 

\section{Discussion}
The first result of this work is that it is possible to start with models with $\mathcal{N}=1$ linear supersymmetry with an unconstrained chiral multiplet and derive the models with non-linearly realized supersymmetry, in a very efficient way. For example, we have reconstructed the complete action \rf{VA} from the model with linear supersymmetry, instead of only the leading terms in the VA goldstino action, as in the procedure in \cite{Komargodski:2009rz}. It is even more surprising that the same procedure for eliminating the heavy sgoldstino works in the case of a local supersymmetry, and leads to an alternative derivation of the pure de Sitter supergravity action \rf{eq.LPuredS}. These results are presented in Sec. 3.

The second result of this paper is the derivation of the explicit pure de Sitter supergravity and the de Sitter supergravity interacting with one matter multiplet, from the general result derived in \cite{Kallosh:2015tea} concerning theories with general coupling, presented there in condensed notation. In the case of pure de Sitter supergravity the explicit action in \rf{eq.LPuredS} is different from the first version derived in \cite{Bergshoeff:2015tra}; the difference corresponds to a different choice of gauge-fixing of the Weyl symmetry, R-symmetry and special supersymmetry in the superconformal version of the theory. Here, following \cite{Kallosh:2015tea}, we used the class of gauges corresponding to a standard unconstrained supergravity, where it is possible to construct de Sitter supergravities with general couplings. These gauges are defined in e.g. \cite{Freedman:2012zz}.
 
An explicit model of de Sitter supergravity interacting with one matter multiplet is presented in \rf{ds+1}. The complete action with all fermions, the gravitino, a fermion from the nilpotent multiplet $\chi^1$ and the inflatino $\chi^2$ has local supersymmetry and de Sitter vacua under the condition
\be
\Lambda = |\rm{f}|^2 + \frac{1}{k_{2\bar{2}}}|\nabla_2\rm{g}|^2 - 3 |\rm{g}|^2 >0.
\ee
as predicted in \cite{Kallosh:2015sea,Kallosh:2015tea}. The gravitino is mixed with the goldstino in this model, which defined the super-Higgs mechanism
\be
\bar{\psi}_\mu\g^{\mu} v +\rm{h.c.},
\ee
where $v= (\chi^1\mathrm{f}+\chi^2\nabla_2\mathrm{g})$. The gravitino eats the goldstino and becomes massive in the unitary gauge $v=0$.
Various choices of gauges for the local supersymmetry can be investigated, starting from the action \rf{ds+1}.
This kind of model is particularly suitable for studies of cosmology and early universe inflation where the matter multiplet $(z^2, \chi^2, F^2)$ is an inflaton supermultiplet, whereas the presence of the nilpotent multiplet $(z^1= {(\chi^1)^2\over 2F^1}, \chi^1, F^1)$ helps to produce inflationary models with a stabilized inflaton partner with the exit to de Sitter vacua of the type studied e.g. in \cite{Carrasco:2015pla}.
The fermionic part of the detailed action here presented will be interesting in these models with regard to reheating, creation of matter, and particle physics.

\section*{Acknowledgements}
We are grateful to E. Bergshoeff, S. Ferrara, D. Freedman, A. Linde, M. Porrati, D. Roest, M. Schillo, J. Thaler, E. van der Woerd, A.~ Van Proeyen and T. Wrase for stimulating discussions. This work is supported by the SITP and by the NSF Grant PHY-1316699. AK is also supported by the K.~A.~Wallenberg Foundation.

\def\jhep{JHEP}
{\small
\bibliography{references.bib}
\bibliographystyle{JHEP1}
}
\end{document}